\documentstyle[12pt]{article}
\begin{document}

\begin{center}

{\Large\bf  Quantum Geometrodynamics for Black Holes and Wormholes}

\vspace{12mm}
{\large V.A.Berezin }
\vspace{3mm}

Institute for Nuclear Research of the Russian Academy of Sciences,

60th October Anniversary Prospect, 7a, 117312, Moscow, Russia

e-mail: berezin@ms2.inr.ac.ru

\vspace{5mm}

{\large A.M.Boyarsky and A.Yu.Neronov}

\vspace{3mm}

Dept.Mathematics and Mechanics, Lomonosov Moscow State Univ.

119899, Moscow, Russia

e-mail: boyarsk@mech.math.msu.su and aneronov@mech.math.msu.su
\end{center}

\newpage

\vspace{7mm}

\begin{center}
{\large\bf  Abstract.}
\end{center}

The geometrodynamics of the spherical gravity with a selfgravitating thin 
dust shell as a source is constructed. The shell Hamiltonian constraint 
is derived and the corresponding Schroedinger equation is obtained. 
This equation appeared to be a finite differences equation. Its solutions 
are required to be analytic functions on the relevant Riemannian surface. 
The method of finding discrete spectra is suggested based on the analytic 
properties of the solutions. The large black hole approximation is considered 
and the discrete spectra for bound states of quantum black holes and wormholes 
are found. They depend on two quantum numbers and are, in fact, 
quasicontinuous.

\newpage
%%%%%%%%%%%%%%%%%%%%%%%%%%%%%%%%%%%%%%%%%%%%%%%%%%%%%%%%%%%%%%%%%%%%%%%%%%%%%
\section{Introduction}
%%%%%%%%%%%%%%%%%%%%%%%%%%%%%%%%%%%%%%%%%%%%%%%%%%%%%%%%%%%%%%%%%%%%%%%%%%%%%
In the absence of the theory of quantum gravity we have to construct 
a new theory every time we want to quantize some classical gravitating 
object. The most interesting models of the kind are black holes and 
cosmological  models . Here we are interested in the quantum black holes 
models. These models deserve consideration for many reasons. The black hole
physics gives us an example of the strong gravitational fields. The existence 
of the event (apparent) horizons causes the Hawking's evaporation of the black
 holes . The fate of the evaporating black holes becomes a subject of 
interest. The quantum theory may throw some light to many problems of the 
classical black hole physics.

There were many  many attempts to construct such quantum models.
The most interesting of them are described in \cite{kuchar}, \cite{thieman},
\cite{filipov}.In these papers the quantum theory of the eternal Schwarzschild 
  black hole was constructed . The authors introduced many useful and 
important  mathematical tools (in the present paper the canonical transformation
found by Kuchar is widely used. but the physical result is rather trivial and 
obvious. Namely the quantum functional depends only on the Schwarzschild 
masses. The reason for this is that the eternal  Schwarzschild black hole
has no dynamical degrees of freedom. That is , all the matter collapsed
classically and all possibly dynamical degrees of freedom died in the 
singularity. In the present paper we treat (or try to ) the simplest
quantum black hole model. ``The simplest'' means that we consider a 
spherically symmetric gravity with a self-gravitating thin dust shell as a 
source. We constructed the classical geometrodynamics for the system.
Quantization of such a model leads to the Schroedinger equation in finite
differences in the coordinate representation. The shift in the argument 
is along imaginary axes which has very important consequences. One of them 
is that the wave function  which are the solution to such an equation 
should be analytical function on the appropriate Riemannian surface.

It should be noted that it is not the first time we are dialing with the
finite differences equation . In the toy model constructed by one of us
the Schroedinger equation in finite differences emerged as a result of the use
 of the proper time quantization. Unlike this  toy model the present 
consideration  deals with the canonical formalism from the very beginning.
Thus , the results do not depend on the choice of time. And the appearance 
of the  finite differences equation is due to the nonlocal nature of the 
corresponding Hamiltonian operator.

In the ordinary quantum mechanics we are dealing with the second order 
differential equations.  And we demand that the solution should be at least 
two times differentiable. To find eigenfunctions and spectrum we need
to specify a class of functions, usually by imposing appropriate boundary
conditions.  In our case of the  finite differences operator we must
specify a class of function by demanding analyticity (except in the branching 
points). Our experience with the toy model shows that the boundary conditions 
help us to select  the wave eigenfunctions
 (though up to the infinite degeneracy)
but the are useless in finding the mass spectrum. To find the spectrum 
we need to know only the analytic properties 
of the solutions, namely their branching points.

It makes our life easier. We do not need to solve the Schroedinger equation.
We should only to investigate the behavior of the solutions in the singular 
points of the corresponding equation.

The plan of the paper is the following. In the Section 2 we remind some facts 
from the classical dynamics of the thin dust shells. The classical 
geometrodynamics is developed in Section 3.  Section 4 is devoted  to the
derivation of the shell's Hamiltonian constraints. The quantum geometrodynamics
of the spherical gravity with the thin shell is considered in Section 5.
Section 6 is dealing with the quasiclassical limit of our Schroedinger
equation which in our case is the same as the large black holes regime.
In this section we found the quantum black hole  and wormhole discrete mass
spectra.

\newpage

%%%%%%%%%%%%%%%%%%%%%%%%%%%%%%%%%%%%%%%%%%%%%%%%%%%%%%%%%%%%%%%%%%%%%%%%%%%%%%
\section{Preliminaries.}
%%%%%%%%%%%%%%%%%%%%%%%%%%%%%%%%%%%%%%%%%%%%%%%%%%%%%%%%%%%%%%%%%%%%%%%%%%%%%%%

The aim of this paper is to consider a geometrodynamics , both classical and quantum of the
spherically symmetric gravitational field with self-gravitating dust thin shell as a source.

We start with the description of the model. This is just a self-gravitating spherically 
symmetric dust thin shell, endowed with a bare mass $M$. The whole space-time is divided into 
three different regions: the inner part ( I ), the outer part ( II ) containing
no matter fields  separated by thin layer III, containing the dust matter of the shell.

The general metric of a spherically symmetric spacetime has the form:
\begin{equation}
\label{metric}
ds^2=- N^2 dt^2 + L^2 ( dr + N^r dt )^2 + R^2 ( d\theta^2 + \sin^2\theta
d\phi^2 )
\end{equation}
where $(t,r,\theta ,\phi )$ are space-time coordinates, $ N, N^r, L, R$ are
some functions of $t$ and $r$ only. Trajectory of the thin shell is some
3-dimensional surface $\Sigma $ in space-time  given by
some function $\hat r(t)$: $\Sigma^3=\{ (t,r,\theta ,\phi ): r=\hat r(t)\} $. In region I $ r< \hat r-\epsilon $,
in a region II $ r>\hat r+\epsilon $, region III is a thin layer $\hat r-\epsilon
<r<\hat r+\epsilon $.

We require that metric coefficients $N,N^r,L$ and $R$ are
continuous functions but jump discontinuities could appear in their derivatives
at the points of $\Sigma $ when the limit $\epsilon\rightarrow 0$ is taken.

Contrast to the flat space-time the normal vector to the surface $R=const$ may not be only spacelike but 
also timelike.In the first case the invariant
\begin{equation} 
F=g^{\alpha\beta}R_{,\alpha}R_{,\beta}>0
\end{equation}
and the corresponding region is called $R$-region (fig. 1)
(here $g_{\alpha\beta}$ is a metric tensor, $g^{\alpha\beta}$ is its inverse,
 $R_{,\alpha}$ denotes the 
partial derivative with respect to the corresponding coordinate, Greek indices run from 0 to 3). 
In the flat case $R$-region occupies the whole space-time. In the second case
\begin{equation}
F<0
\end{equation}
Such a region is called the $T$-region (the notions of $R$- and $T$- regions were introduced in \cite{1}).
It is easy to show that the condition $\dot R=0$ (dot denotes time derivative) cannot be satisfied in a 
$T$-region, hence it should be either  $\dot R>0$ (this region of inevitable 
expansion is called $T_+$-region), or $\dot R<0$ 
(inevitable contraction, a $T_-$-region). Correspondingly, it is impossible to
get $R'=0$ (prime denotes the spatial derivative) in $R$-regions, and a region 
with $R'>0$ is called an 
$R_+$-region, while that with $R'<0$ is an $R_-$ region. The $R_+$- and $R_-$- regions correspond 
to different sides of the Einstein-Rosen bridge (see fig. 1).

The solution of Einstein equations representing the Schwarzschild (spherically symmetric) black hole
is well known and can be put in the form
\begin{equation}
\label{child}
ds^2=-FdT^2+F^{-1}dR^2+R^2(d\theta^2+\sin^2\theta d\phi^2)
\end{equation}
where 
\begin{equation}
F=1-\frac{\displaystyle 2Gm}{\displaystyle R}
\end{equation}
and $m$ is the total mass (energy) of the system, $G$ is the gravitational constant
(equal to the inverse square of Planckian mass, $G=M_{pl}^{-2}$,in the chosen units
with $c=\hbar =1$;
note also that in these units  and the radius 
has dimension of inverse mass).

The metric of 3-dimensional surface $\Sigma^3$ , representing the evolution of 
the thin shell can be written as
\begin{equation}
\left. ds^2\right|_\Sigma =-d\tau^2+\hat R(\tau )(d\theta^2+\sin^2 \theta d\phi^2)
\end{equation}
Here $\tau$ is the proper time of the observer sitting on the shell.
From Einstein equations one obtains the equations of motion of the shell in the 
form
\begin{equation}
\label{Einst}
\sigma_{in}\sqrt{\dot {\hat R}^2+F_{in}}-\sigma_{out}\sqrt{
\dot {\hat R}^2+F_{out}}=\frac{\displaystyle GM}{R}
\end{equation}
( see \cite{shell}). The quantity $\sigma$ has the following meaning, $\sigma =+1$
if radii $R$ increase in the outward normal direction to the shell, and $\sigma =-1$
if radii decrease. Therefore in the $R$-regions $\sigma$ does not change its sign,
the latter being the sign of an $R$-region. It is clear now that on ``our'' side 
of Einstein-Rosen bridge we have $\sigma =+1$, this we shall call the ``black hole case''
, while on the ``other'' side $\sigma =-1$, this we shall call the ``wormhole case''. 
On fig. 2-5 we show all possible junctions of inner and outer regions with 
Schwarzschild masses $m_{in}$ and $m_{out}$ with thin shell of bare mass $M$.
This figures are rather schematic. The space-time inside the shell (to the left
of the trajectories ) depends on whether there are other shells  inside the
given one or not. To avoid such a concretization we draw the part of a complete
inner Schwarzschild space-time with the left infinity. What important here
are the junction of the space-times and the and the positions of the event
 horizons.

\newpage

%%%%%%%%%%%%%%%%%%%%%%%%%%%%%%%%%%%%%%%%%%%%%%%%%%%%%%%%%%%%%%%%%%%%%%%%%
 \section{ Canonical formalism for spherically symmetric gravity with
thin shell.}
%%%%%%%%%%%%%%%%%%%%%%%%%%%%%%%%%%%%%%%%%%%%%%%%%%%%%%%%%%%%%%%%%%%%%%%%%%%

The action functional for the system of spherically symmetric gravitational field and the thin shell is
 \begin{equation}
 \label{action}
 \begin{array}{c}
S=S_{gr}+S_{shell}=\frac{1}{16\pi G}\int\limits_{I+II+III}\
^{(4)}R\sqrt{-g} d^4x + (surface\  terms) - \\ - M\int\limits_{ \Sigma } d\tau
 \end{array}
 \end{equation}
It consists of the standard Einstein-Hilbert action for the gravitational 
field and matter part of the action describes a thin shell of dust. 
The surface terms in the gravitational action and the falloff behavior of 
the metric and
its derivatives were studied in details in \cite{kuchar}. So we will not 
consider this question and will use the results of Kuchar when needed. We will 
be interested in the behavior of the action and constraints on the 
surface $\Sigma^3$ representing the shell's trajectory.   

The complete set of degrees of freedom of our system consists of the set of
$N(r,t), N^r(r,t), L(r,t), R(r,t)$ which describe gravitational field
and $\hat r(t)$ which describes the motion of the shell.

The metric (\ref{metric}) has the standard ADM form for 3+1
decomposition of a space-time with lapse function $N$ , shift vector $N^i=(N^r,0,0)$ and space
metric $h_{ik}=diag( L^2, R^2, R^2\sin^2\theta)$ given foliation of the manifold on
space and time. The scalar curvature density
has the form
  \begin{equation}
 \label{curvature}
 \begin{array}{c}
\ ^{(4)}R\sqrt{-g}=
\\
N\sqrt{h}\left( \ ^{(3)}R+\left( (Tr K)^2-Tr K^2\right)
\right)- \\
- 2\left( \sqrt{h} K)_{,0} \right) + 2 \left( \sqrt{h} KN^i-\sqrt{h}
h^{ij} N_{,j}\right)_{,i}
 \end{array}
 \end{equation}
where $\ ^{(3)}R$ and $K^{ij}$ are the  scalar curvature of a space metric $h_{ij}$
and exterior curvature tensor of a  surface $t=const$. Substituting
expression (\ref{metric}) for the metric into (\ref{curvature}) we obtain
the expression for internal and external curvatures of the surface $t=const$
in the form
 \begin{equation}
\ ^{(3)}R=\frac{
\displaystyle
2}{\displaystyle R^2}\left( 1- \frac{\displaystyle (R')^2}{\displaystyle
L^2}-\frac{\displaystyle 2RR''}{\displaystyle L^2}+\frac{\displaystyle
2RR'L'}{\displaystyle L^3} \right)
\end{equation}
and
\begin{equation}
 \begin{array}{c}
K^i_j=diag(K^r_r, K^\theta_\theta, K^\phi_\phi ) \\
K^r_r=\frac{\displaystyle 1}{\displaystyle NL}\left( \dot L-L'N^r-L(N^r)'
\right), \\
K^\theta_\theta =K^\phi_\phi =\frac{\displaystyle 1}{\displaystyle NR} \left(
\dot R-R'N^r\right)
 \end{array}
 \end{equation}
Here dot and prime denote differentiation in  $t$ and  $r$ respectively.

Contributions to the gravitational action from the terms containing total
derivatives in (\ref{curvature})  give rise to the surface terms which cancel each other at the common boundaries
of regions I, II and II, III. So we are left with the surface terms at 
infinity which were extensively discussed in \cite{kuchar}.
We will turn to them later.

The essential part of the action for gravitational
field is just the ADM part of the action (\ref{action}) with Lagrangian

 \begin{equation}
 \label{ADM}
L_{gr}=\frac{1}{16\pi G}NLR^2\left(\ {(3)}R-(Tr K)^2-Tr K^2\right)
 \end{equation}

Contribution to the  action from the integral over the region III
in the limit $\epsilon\rightarrow 0$ is only due to the term containing second
derivative of $R$,namely 
 \begin{equation}
\int\limits_{III} \frac{1}{16\pi G}NLR^2\ ^{(3)}R =
-\int\limits_{II} \frac{\displaystyle
NRR''}{\displaystyle GL}=
-\int\limits_{\Sigma } \frac{\displaystyle \hat N
\hat R\left[ R'\right] }{
\displaystyle G\hat L}
 \end{equation}
We will denote by hats  variables on $\Sigma $ and by $\left[ {\cal A}\right]
=\lim_{\epsilon\to 0}\left( {\cal A}(\hat r+\epsilon)-{\cal
A}(\hat r-\epsilon)\right)$ a jump of variable ${\cal A}(r)$ on the
shell surface.

Substituting the expression (\ref{metric}) into the shell part of
the action we have:
 \begin{equation}
 \label{matter}
S_{shell}
=
-M\int\limits_\Sigma
\sqrt{ \hat {N}^2-\hat {L}^2 \left(
\hat {N}^r + \dot{\hat {r}}
\right)^2} dt
 \end{equation}

The explicit form of the action (\ref{action}) with metric (\ref{metric})
becomes
  \begin{eqnarray}
 \label{actionone}
 S&=&
\frac{1}{G}\int\limits_{I+II+III}
\left( N
\frac{\displaystyle L}{\displaystyle 2}
\frac{\displaystyle (R')^2}{\displaystyle 2L}-
\left(\frac{\displaystyle RR'}{\displaystyle L}\right) '
+\frac{\displaystyle R}{\displaystyle N}
\left(\dot{R}-R'N^r\right)
\left( (LN^r)'-\dot{L}\right) \right.
\nonumber \\
&+&
\left.
\frac{\displaystyle L}{\displaystyle 2N}
\left( \dot{R}-R'N^r\right)^2
\right)
-\int\limits_{\Sigma }
\left(
\frac{\displaystyle \hat{N}\hat{R}\left[ R'\right] }{\displaystyle \hat{L}}
-m\sqrt{ \hat{N}^2-\hat{L}^2\left( \hat{N}^r+\dot{\hat{r}}\right)^2}
\right) dt
\end{eqnarray}

The canonical formalism for this action can be described in the following way.
Momenta conjugate to corresponding dynamical variables are
 \begin{equation}
 \label{momenta}
 \begin{array}{rcl}
P_N&=&\delta S\left.\right/\delta \dot N=0;\\
P_{N^r}&=&\delta S\left.\right/\delta \dot N^r=0\\
P_L&=&\delta S\left.\right/\delta \dot L =\frac{\displaystyle R}{\displaystyle GN}\left(R'N^r-\dot R\right)\\
P_R&=&\delta S\left.\right/\delta \dot R =\frac{\displaystyle L}{\displaystyle GN}\left( R'N^r-\dot R\right)+
\frac{\displaystyle R}{\displaystyle GN}\left( (LN^r)'-\dot L\right)\\
P_{\hat R}&=&\delta S\left.\right/\displaystyle\delta \dot{ \hat R}=0\\
P_{\hat L}&=&\delta S\left.\right/\delta \dot{\hat L}=0\\
\hat\pi&=&\delta S\left.\right/\delta \dot {\hat r}=\frac{\displaystyle m\hat L^2 ( N^r+\dot{\hat{r}} ) }
{\displaystyle
\sqrt{\hat N^2-\hat L^2 (N^r+\dot{\hat {r}} ) } }
 \end{array}
 \end{equation}
The action (\ref{actionone}) rewritten in the Hamiltonian form becomes
 \begin{equation}
 \label{hamil}
 \begin{array}{c}
S=\int\limits_{I+II}\left( P_L\dot L+P_R\dot R-NH-N^rH_r\right) dr dt+
\int\limits_\Sigma \hat{\pi}\dot{\hat{ r}}- \\
\hat N
\left(
\hat R\left[ R'\right] /(G\hat L) + \sqrt{m^2+ \hat\pi^2/ \hat L^2}
\right) -
\\
\hat N^r \left(-\hat L\left[ P_L\right]-\hat\pi\right) dt
 \end{array}
 \end{equation}
with

 \begin{equation}
 \label{constraints}
 \begin{array}{rcl}
H&=&
G\left( \frac{\displaystyle LP_L^2}{\displaystyle 2R^2}-
\frac{\displaystyle P_LP_R}{\displaystyle R}\right)
+\frac{\displaystyle 1}{\displaystyle G}\left( - \frac{\displaystyle
L}{\displaystyle 2}- \frac{\displaystyle (R')^2}{\displaystyle 2L}+ \left(
\frac{\displaystyle RR'}{\displaystyle L}\right)'\right) \\
H_r&=&P_RR'-LP_L'.  \end{array}
 \end{equation}
where $N, N^r, \hat N$ and $\hat N^r$ are Lagrange multipliers in the 
Hamiltonian formalism.
The system of constraints contain two surface constraints in
addition to usual Hamiltonian and momentum constraints of the ADM formalism.

ADM constraints:
 \begin{equation}
 \label{constraints1}
 \left\{
 \begin{array}{l}
H=0\\
H_r=0\\
 \end{array}
 \right.
 \end{equation}
Shell constraints:
 \begin{equation}
 \label{shellconstraints}
 \left\{
 \begin{array}{l}
\hat H_r=\hat\pi+\hat L\left[ P_L\right] =0\\
\hat H=\frac{\displaystyle R\left[ R'\right] }{\displaystyle GL}+
\sqrt{M^2+\left.\hat\pi^2\right/ L^2}=0
 \end{array}
 \right.
 \end{equation}

\newpage

%%%%%%%%%%%%%%%%%%%%%%%%%%%%%%%%%%%%%%%%%%%%%%%%%%%%%%%%%%%%%%%%%%%%%%%%%%%%%%%%%%%
\section{Kuchar variables.}
%%%%%%%%%%%%%%%%%%%%%%%%%%%%%%%%%%%%%%%%%%%%%%%%%%%%%%%%%%%%%%%%%%%%%%%%%%%%%%%%%%

In the paper \cite{kuchar} Kuchar proposed canonical transformation of 
the variables
$(R, P_R, L, P_L)$ to new canonical set $(R, \bar P_R, M, P_M)$ in
which Hamiltonian and momentum constraints given by (\ref{constraints})
are equivalent to the very simple set of constraints :
 \begin{equation}
 \label{simple}
 \begin{array}{l}
\bar P_R=0\\
M'=0
 \end{array}
 \end{equation}
The idea is to use the Schwarzschild anzatz for the space-time metric 
(\ref{child}) instead of
(\ref{metric}):
 \begin{equation}
 \label{schwar}
ds^2=-F(R,m) dT^2+\frac{\displaystyle 1}{\displaystyle F(R,m)} dR^2+
R^2 (d\theta^2+\sin^2\theta d\phi^2)
 \end{equation}
where $T, R$ and $m$ are some functions of $(r,t)$ and $F(R,m)=1-\left. 2Gm
\right/ R$, and m, in general, is a function of r, m=m(r).
 Equating the two forms of the metric (\ref{metric}) and
(\ref{schwar}) we obtain the transformation between the two sets of
dynamical variables.

The explicit form of the transformation is
 \begin{equation}
 \label{transformation}
 \begin{array}{rcl}
L&=&\sqrt{ \frac{\displaystyle (R')^2}{\displaystyle F}-FP_m^2}\\
P_L&=&\frac{\displaystyle RFP_m}{\displaystyle G}\sqrt{\frac{\displaystyle
(R')^2}{\displaystyle F}-FP_m^2}\\ R&=&R\\ \bar
P_R&=&P_R+\frac{\displaystyle P_m}{\displaystyle 2G}+ \frac{\displaystyle
FP_m}{\displaystyle 2G}+ \frac{\displaystyle
(RFP_m)'RR'-RFP_m(RR')'}{\displaystyle GRF\left( (R')^2\left.  \right/
F-FP_m^2\right) }
 \end{array}
 \end{equation}
where $P_m=-T'$.

The Liouville form
 \begin{equation}
 \label{liuvil}
\Theta=\int P_R \dot R+P_L \dot L
 \end{equation}
can be expressed in the new variables as follows:
 \begin{eqnarray}
 \label{important}
\Theta&=&\int P_m\dot m +\bar P_R\dot R+\frac{\displaystyle \partial}{
\displaystyle \partial t}\left( LP_L+
\frac{\displaystyle 1}{\displaystyle 2G}RR'\ln{\left|
\frac{\displaystyle RR'-LP_LG}{\displaystyle RR'+LP_LG}\right|}\right)
\nonumber \\
&+&
\frac{\displaystyle \partial}{\displaystyle \partial r}\left(
\frac{\displaystyle 1}{\displaystyle 2G}R\dot{ R}\ln{\left|
\frac{\displaystyle RR'+LP_LG}{\displaystyle RR'-LP_LG}\right| }\right).
 \end{eqnarray}
When there is no shell the  total derivatives in (\ref{important}) give 
rise to some surface terms at infinities.
 As shown by Kuchar \cite{kuchar} the appropriate falloff conditions
at infinities make the last  surface  term from (\ref{important}) zero. Then it follows from
(\ref{important}) that  $(R, \bar P_R, m, P_m)$ form a canonical set of 
variables and equation (\ref{transformation}) 
describes a canonical transformation between   $(R, P_R, L, P_L)$ and 
 $(R, \bar P_R, m, P_m)$ .

\newpage

%%%%%%%%%%%%%%%%%%%%%%%%%%%%%%%%%%%%%%%%%%%%%%%%%%%%%%%%%%%%%%%%%%%%%%%%%%%
\section{Canonical variables and Hamiltonian constraint on the shell.}
%%%%%%%%%%%%%%%%%%%%%%%%%%%%%%%%%%%%%%%%%%%%%%%%%%%%%%%%%%%%%%%%%%%%%%%%%%%%%%

%%%%%%%%%%%%%%%%%%%%%%%%%%%%%%%%%%%%%%%%%%%%%%%%%%%%%%%%%%%%%%%%%%%
\subsection{Shell variables.}
%%%%%%%%%%%%%%%%%%%%%%%%%%%%%%%%%%%%%%%%%%%%%%%%%%%%%%%%%%%%%%%%%%%%%%%%%%%%%%%

In the presence of the thin shell the situation is different. Surface terms now should not be
neglected.

Let us do the transformation (\ref{transformation}) in regions I and II of
our space-time.The Liouville form of our Hamiltonian system (\ref{hamil})
has the form
 \begin{equation}
\tilde \Theta =\int\limits_{I+II} P_R\dot R+P_L\dot L+\int\limits_\Sigma
\hat\pi\dot{\hat r}.
 \end{equation}
After integration the total derivatives in (\ref{important}) give some contribution to
the Liouville form on $\Sigma$:
 \begin{eqnarray}
\tilde \Theta &=&\int\limits_{I+II} \bar P_R\dot R+P_m\dot m+
\int\limits_\Sigma
\left[LP_L+
\frac{\displaystyle 1}{\displaystyle 2G}
RR'\ln{\left|
\frac{\displaystyle RR'-LP_LG}{\displaystyle RR'+LP_LG}\right|} \right]
\dot{\hat r}
dt
\nonumber \\
&-&\int\limits_\Sigma \left[
\frac{\displaystyle 1}{\displaystyle 2G}R\dot R\ln{\left|
\frac{\displaystyle RR'+LP_LG}{\displaystyle RR'-LP_LG}\right| }\right]+
\int\limits_\Sigma \hat\pi\dot{\hat r}
\nonumber \\
&=&
\int\limits_{I+II}P_m\dot m+\bar P_R\dot R+\int\limits_\Sigma \hat p\dot{\hat r}
+\int\limits_\Sigma \hat P_{\hat R} \dot{\hat R}
 \end{eqnarray}
where we denoted
 \begin{equation}
 \label{newtransf}
 \begin{array}{rcl}
\hat p&=&\hat\pi+L\left[ P_L\right]\\
\hat P_{\hat R}&=& \left[ \frac{\displaystyle 1}{\displaystyle 2G}
R\ln{\left|
\frac{\displaystyle RR'-GLP_L}{\displaystyle RR'+GLP_L}\right|} \right]
 \end{array}
 \end{equation}
and made use of the identity
\begin{equation}
\dot{\hat R}=
\frac{\displaystyle d}{\displaystyle dt} R(t,\hat r(t))=
(\dot R(t,r)+R'(t,r)\dot{\hat r}(t))\left.\right|_{r=\hat r(t)}
\end{equation}
We see that this canonical transformation involves all the set of coordinates
in the phase space $\Pi=\{  (R(r,t), P_R(r,t), L(r,t), P_L(r,t), \hat {r}(t),
\hat{\pi}(t)) \}$
 according to the formulae (\ref{transformation}) and (\ref{newtransf}).
Moreover it introduces additional pair of canonically conjugate variables
$(\hat R, \hat P_{\hat R})$ on the shell.

In both inner and outer regions I and II constraints are simplified due to the
canonical transformation as it was in the absence of the shell (\ref{simple}).
The surface momentum constraint $\hat H_r=0$ (\ref{constraints}) takes the form
 \begin{equation}
\label{exclude}
\hat p=0
 \end{equation}

\newpage
%%%%%%%%%%%%%%%%%%%%%%%%%%%%%%%%%%%%%%%%%%%%%%%%%%%%%%%%%%%%%%%
\subsection{Shell constraint. Special case.}
%%%%%%%%%%%%%%%%%%%%%%%%%%%%%%%%%%%%%%%%%%%%%%%%%%%%%%%%%%%%%%

In order to investigate the form of Hamiltonian surface constraint in
new variables let us choose coordinates $(r,t)$ in a specific way.
If coordinate lines $t=const$ cross the shell perpendicularly then
 \begin{equation}
\dot{\hat r} =-\hat N^r
 \end{equation}
This means according to (\ref{momenta}) that
 \begin{equation}
\hat\pi=0
 \end{equation}
Then the surface momentum constraint gives
 \begin{equation}
GL\left[ P_L\right] =R\left[ FP_M\right] =0
 \end{equation}

Let us denote 
\begin{equation}
\label{27}
\gamma=\left. FP_M\right/L
\end{equation}
We have from (\ref{newtransf})
 \begin{equation}
 \label{helpful'}
\hat {P}_{\hat R}=
\frac{\displaystyle 1}{\displaystyle 2G}
R\left[ \ln\left|
\frac{\displaystyle R'-\gamma }{\displaystyle R'+\gamma }\right|\right] =
\frac{\displaystyle R}{\displaystyle G}\ln \left(\left|
\frac{\displaystyle R'_{in}+\gamma }{\displaystyle R'_{out}+\gamma }\right|
\sqrt{\left|\frac{\displaystyle F_{out}}{\displaystyle F_{in}}\right|}\right)
 \end{equation}
From (\ref{transformation} we get 
 \begin{equation}
 \label{helpful}
\frac{\displaystyle R'}{\displaystyle L}=\sigma\sqrt{F+\gamma^2}
 \end{equation}
where $\sigma=\pm 1$ is the sign function 
taking its values according to whether radii increase in the outward normal direction 
to the shell ( $\sigma =+1$ ) or they  decrease ( $\sigma =-1$ ).
Using (\ref{helpful'}) and (\ref{helpful}) we could find the expression for  $\gamma$.
 Then from (\ref{helpful}) we could determine the jump of $R'$ across
the shell surface as a function of $\hat P_{\hat R}, \hat R, m_{in}$ and $m_{out}$. Substituting this  
into the surface Hamiltonian constraint (\ref{shellconstraints}) we obtain the following expression
\begin{equation}
\label{hamilton}
\hat H=
\frac{\displaystyle \sigma_{in} R}{\displaystyle G}
\sqrt{\sqrt{F_{out} }-\sqrt{F_{in}}  \exp \left( \frac{\displaystyle G\hat P_{\hat R}}{\displaystyle R}
\right)}
\sqrt{\sqrt{F_{out}}-\sqrt{F_{in}}  \exp \left( -\frac{\displaystyle G\hat P_{\hat R}}{\displaystyle R}
\right)} - M=0
\end{equation}

\newcommand{\R}{\dot{\hat R}}
\newcommand{\dfrac}[2]{\displaystyle\frac{#1}{#2}}
\newcommand{\text}[1]{\mbox{#1}}

\newpage
%%%%%%%%%%%%%%%%%%%%%%%%%%%%%%%%%%%%%%%%%%%%%%%%%%%%%%%%%%%%%%%%%%%%%%%%%%%
\subsection{Shell constraint. General case.}
%%%%%%%%%%%%%%%%%%%%%%%%%%%%%%%%%%%%%%%%%%%%%%%%%%%%%%%%%%%%%%%%%%%%%%%%%%%%%%

In the previous subsection we managed to derive the shell constraint by finding so
some continuous variable, namely $\gamma$ (Eqn.\ref{27}).Now we will try to do
 the same trick in general case $\pi\neq 0$. 
Let's consider the full time derivative of the shell radius.
\begin{equation}
\label{DotHatR}
\R\equiv \frac d{dt}\hat R\left( r\left( t\right)
,t\right) =\dot R+R^{\prime }\dot r
\end{equation}
 Using the definition $P_L=-\frac
R{GN}\left( \dot R-R^{\prime }N^r\right) $ we get:
\begin{equation}
\label{dotR}
\dot {\hat R}=-\frac{GNP_L}R+R^{\prime }\left( \dot r+N^r\right)
\end{equation}
Remembering that
$$
\pi \equiv \frac{ML^2\left( N^r+\dot r\right) }{\sqrt{N^2-L^2\left( N^r+\dot
r\right) }} 
$$
we can find
$$
L\left( N^r+\dot r\right) =\frac{\pi N}{L\sqrt{M^2+
\dfrac{\pi ^2}{L^2}}}
$$
Eqn.( \ref{dotR}) now reads

\begin{equation}
\frac{\dot {\hat R}}N=-\frac{P_L}R+\frac{%
R^{\prime }}{L}\frac \pi {L\sqrt{M^2+\dfrac{\pi ^2}{L^2}}}
\end{equation}
It is now easy to see that the jump of $\R$ across the shell is a linear
combination of constraints
\begin{equation}
\left[ \dot {\hat R}\right] =\frac {GN}R \left( \chi H -\frac{H^r}L\right)
\end{equation}
where
$$
\chi =\frac \pi {L\sqrt{M^2+\dfrac{\pi ^2}{L^2}}}
$$
To go further  we need Eqn.(\ref{}) which we now rewrite as follows
\begin{equation}
\label{beta}
\beta \equiv {\large e}^{\frac{\displaystyle G\hat{P}_R}{\displaystyle R}}=
\frac{\left( \frac{\R}N+G\frac{P_L}R\left( 1+\chi
\right) \right) _{in}}{\left( \frac{\R}N+G\frac{P_L}R\left( 1+\chi
\right) \right) _{out}}\equiv \frac{\alpha +y_{in}\left( 1+\chi \right) }{%
\alpha +y_{out}\left( 1+\chi \right) }
\end{equation}
where $\alpha =\frac{\R}N$ and $y=G\frac{P_L}R$.\\ 
The next step is to find  the relation between $\alpha$ and $\R$. From the 
definitions of $P_L$ and $\R$ we have
$$
\frac {{R^{\prime }}^2}{L^2}=F+y^2.=\frac \alpha \chi +\frac y\chi
$$.
Solving this  for $y$ we get
\begin{equation}
\label{y}
y=\frac{\alpha \pm \sqrt{\alpha ^2-\left( \chi ^2-1\right) \left( F\chi
^2-\alpha ^2\right) }}{\chi ^2-1}
\end{equation}

Substituting this expression back into Eqn. \ref{beta} we obtain 
after simple algebraic transformations 
\begin{eqnarray}
&\beta& =\frac{z-\sigma _{in}\sqrt{z^2+F_{in}}}{z-\sigma _{out}
\sqrt{z^2+F_{out}}}\\
where &z^2&=\frac{\alpha ^2}{1-\chi ^2}
\end{eqnarray}
Making use of the momentum constraint we  can wright the jump of $y$ as follows 

\begin{equation}
\begin{array}{c}

\sigma _{out}\chi
\sqrt{\alpha ^2+\left( 1-\chi ^2\right) F_{out}}=\sigma _{in}\chi \sqrt{%
\alpha ^2+\left( 1-\chi ^2\right) F_{in}}-G\frac MR\chi \sqrt{1-\chi ^2}%
  \\  \\
\end{array}
\end{equation}
Note that this is just the Eqn. (\ref{Einst}) if we choose proper time
(substituting $\dot {\hat R}^2$for $z^2$.) \\
Squaring this equation  we get
\begin{equation}
\left\{
\begin{array}{c}
\sigma _{in}
\sqrt{z^2+F_{in}}=-\frac{R\left[ F\right] }{2MG}+\frac {MG}{2R} \\  \\
\sigma _{out}\sqrt{z^2+F_{out}}=-\frac{R\left[ F\right] }{2MG}-
\frac {MG}{2R}
\end{array}
\right.
\end{equation}
\begin{equation}
z=\pm\sqrt{\left( \frac{R\left[ F\right] }{2MG}\right) ^2-\frac 12\left(
F_{out}+F_{in}\right) +\frac{M^2G^2}{4R^2}}
\end{equation}
Finally we get for the shell constraint
\begin{equation}
\label{frac}
\beta - \frac{z+\dfrac{R\left[ F\right] }{2MG}-\dfrac {MG}{2R}}{z+\dfrac{R\left[
F\right] }{2MG}+\dfrac {MG}{2R}}=0
\end{equation}
This constraint can be rewritten in a more elegant way
 \begin{equation}
\label{main}
\begin{array}{c}
\sqrt{F_{out}}\left(\pm\sqrt{ \left( \frac{[m]}M \right)^2-1+
G\frac {m_{out}+m_{in}}R+\frac {M^2G^2}{4R^2} } -
\frac{[m]}M+\frac{MG}{2R} \right)e^{ \frac {G\hat{P}_R}{2R} }-\\
\sqrt{F_{in}}\left(\pm\sqrt{\left( \frac{[m]}M \right)^2-1+
G\frac{m_{out}+m_{in}}R+\frac{M^2G^2}{4R^2}}-
\frac{[m]}M-\frac{MG}{2R} \right) e^{ - \frac {G\hat{P}_R}{2R}} =0
\end{array}
\end{equation}
where the Schwarzschild anzatz was substituted for $F$'s .\\

It can be easily shown that the above constraint is equivalent to that 
one derived in the previous subsection. And this proves that 
the Hamiltonian constraint (\ref{hamilton}) is valid for nonzero values of $\pi$ as well.
We see that the only remained classical constraint of the shell Hamiltonian 
dynamics can be written in various equivalent forms (e.g. , Eqns (\ref{hamilton}),
(\ref{frac}) or (\ref{main}). But, of course, quantum mechanically all
these forms are no more equivalent. So , we need some criteria to choose among
 them. One of this criteria well be considered in the last section.
In what follows we will consider the following squared version of the 
Hamiltonian constraint (\ref{hamilton}) as the suitable classical counterpart
for the quantum constraint for the wave function $\Psi$
\begin{equation} 
\label{Cons}
C=F_{out}+F_{in}-\sqrt{F_{out}}\sqrt{F_{in}}\left( \exp\frac {G\dot P_R} {R}+
 \exp - \frac {G\dot P_R} {R}\right) - \frac{M^2G^2}{R^2}
\end{equation}

 The Hamiltonian constraint (\ref{hamilton}) was derived under the assumption 
that both $F_{in}$ and $F_{out}$
are positive . It is possible, of course to derive analogous constraints in $T_\pm$-regions, 
where $F<0$. But, instead, we make the following substitution
\begin{equation}
\sqrt{F}\rightarrow F^{1/2}
\end{equation}
and consider this function as a function of complex variable. Then the point of the horizon
$F=0$ becomes a branching point , and we need the rules of the bypass. We assume the following 
\begin{equation}
\label{bypass2}
\begin{array}{rl}
F^{1/2}&=\left| F\right| e^{i\phi}\\
\phi =0& \mbox{\ in\ } R_+\mbox{-region}\\
\phi = \pi /2& \mbox{\ in\ } T_-\mbox{-region}\\
\phi =\pi & \mbox{\ in\ } R_-\mbox{-region}\\
\phi =-\pi /2& \mbox{\ in\ } T_+\mbox{-region}\\  
\end{array}
\end{equation}
for the black hole case , and 
\begin{equation}
\label{bypass3}
\begin{array}{rl}
\phi =\pi& \mbox{\ in\ } R_+\mbox{-region}\\
\phi = -\pi /2& \mbox{\ in\ } T_-\mbox{-region}\\
\phi =0 & \mbox{\ in\ } R_-\mbox{-region}\\
\phi =\pi /2& \mbox{\ in\ } T_+\mbox{-region}\\  
\end{array}
\end{equation}
for the wormhole case.
The reason for  such analytical continuation is that we are able to get the single
equation on the wave function $\Psi$ which covers all four patches of the complete
Penrose diagram for the Schwarzschild spacetime.Some consequences of this fact
will become evident in section 6.

\newpage 
%%%%%%%%%%%%%%%%%%%%%%%%%%%%%%%%%%%%%%%%%%%%%%%%%%%%%%%%%%%%%%%%%%%%%%%%%%%%%%%%%%%
\section{Quantized spherical gravity with thin shell.}
%%%%%%%%%%%%%%%%%%%%%%%%%%%%%%%%%%%%%%%%%%%%%%%%%%%%%%%%%%%%%%%%%%%%%%%%%%%%%%%%%%

We now turn to the Dirac constraint quantization procedure . 

It is convenient to make a canonical transformation from
 $(\hat R, \hat P_{\hat R})$
to $(\hat S, \hat P_S)$:
\begin{equation}
\label{area}
\left\{
\begin{array}{rcl}
\hat S&=&\frac{\displaystyle \hat R^2}{\displaystyle (2GM)^2}=\frac{\displaystyle \hat R^2}{\displaystyle R_g^2}\\
\hat P_S&=&R_g^2\frac{\displaystyle \hat P_{\hat R}}{\displaystyle 2R}
\end{array}
\right.
\end{equation}
where $R_g$ is the gravitational radius of the shell. Dimensionless variable $\hat S$ is the surface area of 
the shell measured in the units of horizon  area of the shell of mass $M$.  

The phase space of our model consist of coordinates 
$(R(r), \tilde P_R(r), m(r),\\
 P_m(r), \hat S, \hat P_S, \hat r, \hat p_r)$
$r\in (-\infty, \hat r-\epsilon )\bigcup (\hat r+\epsilon, \infty)$.
Then the wave function in coordinate representation depends on configuration space coordinates:
\begin{equation}
\label{wave}
\Psi=\Psi(R(r), m(r), \hat S, \hat r)
\end{equation}
and all the momenta become operators of the form
\begin{equation}
\label{operators}
\begin{array}{cc}
\tilde P_R(r)=-i\left.\delta\right/\delta R(r)& P_m(r)=-i\left.\delta\right/
\delta m(r)\\
\hat P_S=-i\left.\partial\right/\partial \hat S& \hat p_r=-i\left.\partial\right/\partial \hat r
\end{array}
\end{equation}
ADM and shell constraints (\ref{constraints1}) and (\ref{shellconstraints}) become operator equations on
$\Psi$. The set of ADM constraints is equivalent to the set of constraints (\ref{simple}) in Kuchar variables
which could be easily solved on quantum level. Indeed, in the regions I and II the equations
\begin{equation}
\label{qkuchar}
\left\{
\begin{array}{rcl}
\left.\partial \Psi\right/\partial R(r)&=&0\\
M'(r)\Psi&=&0
\end{array}
\right.
\end{equation}
express the fact that wave function does not depend on $R(r)$ and the dependence on $M(r)$ is
reduced in each region I and II to $\Psi\equiv \delta(M-M_{\pm})$ where $M_{\pm}$ defined
in the regions I (-) and II (+) do not depend on $r$. They equal to Schwarzschild masses 
in the inner and outer regions $M_{in}$ and $M_{out}$ in (\ref{main}).

The set of shell constraints (\ref{shellconstraints}) impose further restrictions on $\Psi$. 
First of them takes the form
\begin{equation}
\left.\partial \Psi\right/\partial \hat r=0
\end{equation}
 in new variables according to (\ref{exclude}). So the only nontrivial equation is the shell
constraint (\ref{Cons}) (or (\ref{hamilton},(\ref{frac}),(\ref{main})which
are classically equivalent to the Eqn.(\ref{Cons}))
\begin{eqnarray}
\hat  C \left(m_+,m_-,\hat S, -i\hbar \partial /\partial \hat S\right)=0 \\
\nonumber
\Psi = \Psi (m_+,m_-,\hat S)
\end{eqnarray}
The operator $\hat C$ contains the exponent of the of the momentum $\hat P_S$. 
This exponent becomes an operator  of finite displacement when $\hat P_S$
becomes differential operator:
\begin{equation}
e^{\frac{G \hat P_R}{R}}=e^{\frac { \hat P_S}{2GM^2}}\Psi=
e^{-i\frac{m^2_pl}{M^2}\frac{\partial}{\partial \hat S}}\Psi =
\Psi(m_+,m_-,\hat S -\xi i)
\end{equation}
where $m_{pl}$ is Plank mass and $\xi =\frac {1}{2} (\frac {m_{pl}}{ M})^2 $ 

The constraint $C$ is nonlinear in $\hat S$ and $ P_S$ so the question of
 ordering should be solved when replacing dynamical variables by operators. 
It is proposed in \cite{berezin}to choose the symmetric ordering
\begin{equation}
\label{order}
A(\hat S)B(\hat P_S)\rightarrow \left. 1\right/ 2 \left\{  A(\hat S) B(-i\hbar\partial\left.\right/\partial\hat S)+
\overline{B}(i\hbar\partial\left.\right/\partial\hat S)\overline{A}(\hat S)\right\}
\end{equation}
where $A$ and $B$ are some functions of $\hat S$ and $\hat P_S$ respectively and
 $\overline{A}$ denotes complex conjugation.

With operator ordering (\ref{order}) we must add to the constraint (\ref{Cons})
 the complex conjugate part .

The constraint $\hat C$ becomes an equation in finite differences  if we 
express $\hat R$ through $\hat S$ and substitute the expression (\ref{finite})
 to  the differential  operator. Finally we get

\begin{equation}
\label{main'}
\begin{array}{c}
F_{out}^{1/2}F_{in}^{1/2}\left( \Psi (s+ i\xi)+\Psi (s- i\xi)\right)+
\overline{F_{out}^{1/2}F_{in}^{1/2}}(s+ i\xi)\Psi (s+ i\xi)+\\
\overline{F_{out}^{1/2}F_{in}^{1/2}}(s- i\xi)\Psi (s- i\xi)=
2(F_{out}+F_{in}-\frac{1}{\displaystyle {4s}})\Psi(s)
\end{array}
\end{equation}

We have mentioned already  that the classically equivalent constraints  give
inequivalent quantum theories. This is well known fact. We suggest that the 
criterion to choose the correct  quantum theory is the  behavior of the wave 
functions in the quasiclassical regime.In our case this means the large black 
holes limit. Indeed , the parameter $\zeta =\frac {1}{2} (\frac {m_{pl}}{ m})^2 $
becomes small for large masses , and the expansion with respect to this 
parameter is equivalent to the expansion in Planckian constant $\hbar$ 
($m_{pl}=\sqrt{\hbar / Gc}$). In the next Section we will consider this
quasiclassical limit and show that our choice for the quantum constraint 
is a good one (at least in the case of one thin shell).
At the and of this Section we would like to make an important remark.
Our quantum equation \ref{main'} (which is just a Schroedinger equation )
is the equation in finite differences rather than differential equation,
and the shift in argument is along an imaginary axis. In the case of 
differential equation we we require the solution to be differentiable 
sufficiently many times . Similarly, we have to demand the solutions of
our finite differences equation \ref{main'} to be analytical functions.
This condition is very restrictive but unavoidable. Our previous experience
(see \cite{shell}) shows that it is the analyticity of the wave functions 
and not the boundary conditions that lead to the existence of the discrete
mass (energy) spectrum for bound states. How it works in the quasiclassical 
regime we will see in the next Section. 

\newpage
%%%%%%%%%%%%%%%%%%%%%%%%%%%%%%%%%%%%%%%%%%%%%%%%%%%%%%%%%%%%%%%%%%%%%%%%%%%%%%
\section{Large black holes.}
%%%%%%%%%%%%%%%%%%%%%%%%%%%%%%%%%%%%%%%%%%%%%%%%%%%%%%%%%%%%%%%%%%%%%%%%%%%%%%

 The finite difference equation (\ref{main'}) becomes an ordinary differential equation in quasiclassical
limit which is the same as the limit of large ( $m\gg m_{pl}$ ) black holes. Indeed the parameter of
finite displacement of the argument of $\Psi$ in (\ref{main'})
  $\xi =\left( m_{pl}\left.\right/ M\right) $ 
becomes small and we could cut the Tailor  expansion
\begin{equation}
\Psi (\hat S+\xi i)=\Psi (\hat S)+i\xi \Psi'(\hat S)-
\frac{\displaystyle \xi^2}{\displaystyle 2}\Psi''(\hat S)+...  
\end{equation}
at the second order.

Now we will analyze the behavior of the solutions of equation (\ref{main'}) in
 this quasiclassical limit at singular points.

We will restrict the consideration to the case of flat inner region $m_{-}=0$,
 so we denote $m_{+}=m$. 
It is convenient to redefine the area variable $\hat S$ so that
\begin{equation}
\left\{
\begin{array}{rcl}
S&=&\frac{\displaystyle \hat R^2}{\displaystyle (2Gm)^2}=\frac{\displaystyle \hat R^2}{\displaystyle \tilde R_g^2}\\
P_S&=&\tilde R_g^2\frac{\displaystyle \hat P_{\hat R}}{\displaystyle 2R}
\end{array}
\right.
\end{equation}
area is now measures in the units of horizon area of a black hole with Schwarzschild mass $M$ ($\tilde R_g$ is its gravitational
radius). In these units the displacement parameter
\begin{equation}
\zeta =\frac{\displaystyle m_{pl}}{\displaystyle m}
\end{equation}
and the equation (\ref{main'}) reads as
\begin{equation}
\label{main''}
\begin{array}{c}
e^{i\phi }\sqrt{|F|}\left( \Psi (s+ i\zeta)+\Psi (s- i\zeta)\right)+
e^{-i\phi }\sqrt{|F|}(s+ i\zeta)\Psi (s+ i\zeta)+\\
e^{-i\phi }\sqrt{|F|}(s- i\zeta)\Psi (s- i\zeta)=
2(F-\frac{1}{\displaystyle {4s}})\Psi(s)
\end{array}
\end{equation}

where $\phi$ is the phase of $F^{1/2}$ . It should be chosen in different $R$- and $T$-
regions according to the arguments of section 5. In the last formula we must take the Tailor expansion on $\xi$ up 
to second order. 
\begin{equation}
\label{expansion}
\begin{array}{rcl}
\left. \Psi\right|_{S\pm \zeta i}&\approx&\Psi(S)\pm \Psi'(S)\zeta i-
\frac{\displaystyle \zeta^2}{\displaystyle 2}\Psi''(S)\dots
\\
\\

\left. F^{\frac{1}2}\right|_{S\pm\zeta i}
&=&\sqrt{1-\frac{\displaystyle 1}{\displaystyle \sqrt{s\pm \zeta i}}}\approx 
F^{\frac{1}2}
\left( 
1\pm
\frac{\displaystyle 1}{\displaystyle 2FS^{3/2}}\zeta i+
\left(
\frac{\displaystyle 3}{\displaystyle 8FS^{5/2}}+
\frac{\displaystyle 1}{\displaystyle 8F^2S^3}
\right)
\zeta^2
\right) \dots\\ 
\\
 
\end{array}
\end{equation} 
This leads to ordinary differential equations of second order, which are 
different in $ R_+, R_-, T_+ $ and $T_-$ regions
due to the different values of the phases  in Eqn.(\ref{main''}). 
The interesting for us singular points of these differential equations are
\begin{equation}
S=\infty  \mbox{\ and\ } S=1.
\end{equation}
In the quasiclassical limit our requirement of the analyticity of the solutions
 to  the exact equation (\ref{main''}) transforms into the requirement that the
branching points of the leading terms in the solutions to the approximate
equations should be of the same kind . Thus, we need to keep only those terms
in the corresponding equations that give us these leading terms. Below we 
consider  the black hole case only. The results are easily translated to the 
wormhole case.

The singular point $S=\infty$ in the region $R_+$ lies in a classically 
forbidden region as far as we restrict ourselves with bound motions of 
the shell only. In order to analyze the behavior of $\Psi$ in this region
we should take (\ref{main''}) with $\phi =0$ and expand all the quantities 
in terms of $y$, where $s=(1+y)^2$
The result is   
\begin{equation}
\label{infty}
\Psi_{yy}-\frac{1}y \Psi_{y}+ \frac{1}{\displaystyle \zeta^2}
\left( 1-\frac{\displaystyle M^2}{\displaystyle m^2}+
\frac{1}{2y}(2-\frac{\displaystyle M^2}{\displaystyle m^2})\right)\Psi=0
\end{equation}
The leading term of the solution is
\begin{eqnarray}
\Psi &\sim &
y^{\displaystyle \frac{1}{2}-\frac
{\displaystyle \frac{\displaystyle M^2}{\displaystyle m^2}-2}
{\displaystyle 4\mu\zeta^2}}
\exp(-\mu y), \\
\nonumber
\mu&=&\frac{1}{\zeta} 
\sqrt{\frac{\displaystyle M^2}{\displaystyle m^2}-1},
\ y\gg\zeta
\end{eqnarray}

For another singular point in $R_+$ region , that is for $S\rightarrow 1+0$
we have $(s=(1+z^2)^2)$
\begin{equation}
\label{z}
\Psi_{zz}- 3z\Psi_{z}+ \frac{16z}{\displaystyle \zeta^2}
\left( 1-\frac{\displaystyle M^2}{\displaystyle 4m^2}\right)\Psi=0
\end{equation}
with leading term 
\begin {eqnarray}
\Psi \sim 
1-\frac{8}{\displaystyle 3\zeta^2}
\left( 1-\frac{\displaystyle M^2}{\displaystyle 4m^2}\right)y^{3/2}\\
\nonumber
y=\sqrt{z}, \ s\gg\zeta,\ y\gg\zeta,\ \ \zeta\ll 1
\end{eqnarray}

Comparing the types of the branching points at $s\rightarrow\infty$
and $s\rightarrow1+0$ we can conclude that 
\begin{equation}
\label{n}
\frac
{\displaystyle 2-\frac{\displaystyle M^2}{\displaystyle m^2}}
{\displaystyle 4\zeta\sqrt{\frac{\displaystyle M^2}{\displaystyle m^2}-1}}
=n, \ \ n=integer
\end{equation}
This is the first quantization condition. We will not consider here the 
wormhole case . Note only that , as can be shown , the positive values 
of quantum number n correspond to black holes while negative n correspond 
to wormholes.

In the $T_-$-region (which  classically is a region of inevitable contraction)
, i.e., for $s\rightarrow1-0$ ($s=(1+y)^2,\ y<0,\ \zeta\ll|y|\ll1  $)
have first order differential equation (due to complex conjugation introduced
earlier the leading terms containing second derivatives of the wave function
cancel each other) with following leading term in the solution
\begin {equation}
\Psi \sim 
\exp\left(i\frac{8}{\displaystyle 3\zeta^2}
\left( 1-\frac{\displaystyle M^2}{\displaystyle 4m^2}\right)(-y)^{3/2}\right)
\end{equation}
which is just the ingoing wave as it should be expected for the quasiclassical
limit in the region of the inevitable contraction.That is why we have chosen 
the of the function $(F)^{1/2}=e^{i\phi}|F|^{1/2}$ in the $T_-$region
(Which classically  the region of inevitable expansion) the choice of 
$\phi=-\frac {\pi }2$ leads to the outgoing wave as a solution.
Note also that to the our requirement of the analyticity the solution 
(leading term) in the $T_-$region should be the analytical continuation of the
solution in the $R_+$region. And we see that this is indeed the case.

We do not consider here separately the asymptotics  in $R_-$ region near the
horizon $(s\rightarrow1+0)$ because it differs from the corresponding solution 
in $R_+ $-region only by the sign in front of the second term.

Let us now turn to the asymptotics of the solutions in $R_-$-region for 
$s\rightarrow\infty$. Due to the minus sign in front of $F^{1/2}$ the 
equation for the wave function in a $R_-$-region is quite different from
that in a $R_+$-region 

\begin{equation}
\label{yy}
\Psi_{yy}-\frac{1}y \Psi_{y}- \frac{1}{\displaystyle \zeta^2}
\left( 16y^2+1-\frac{\displaystyle M^2}{\displaystyle m^2})\right)\Psi=0
\end{equation}
The leading term of the asymptotic is now the following 
\begin{equation}
\Psi \sim 
y^{\displaystyle \frac
{\displaystyle \frac{\displaystyle M^2}{\displaystyle m^2}-1}
{\displaystyle 8 \zeta}}
\exp -(\frac {2}\zeta y^2)
 \end {equation}
Note that the falloff in the $R_-$-region is much faster than it is in the 
$R_+$-region. This is a quite reasonable  result because it means that the
 quantum shell in the black hole case can penetrate into the $R_-$-region 
(which is completely forbidden for the classical motion) but the probability
 of such  an event is negligible small.

And , again, comparing the types of the branching points at $s\rightarrow1+0$
and $s\rightarrow\infty$ in the $R_-$-region we get 
   
\begin{equation}
\label{p}
{\displaystyle \frac
{\displaystyle \frac{\displaystyle M^2}{\displaystyle m^2}-1}
{\displaystyle 8 \zeta}}=\frac{1}2+p, \qquad p=positive \quad integer
 \end {equation}
The appearance of the4 second quantum number is rather surprising  but it
allows some explanation. We discuss this point in the last Section.

Combining (\ref{n}) and (\ref{p}) we get
\begin{equation}
\label{spectr}
{\displaystyle \frac
{\displaystyle (\frac{\displaystyle M^2}{\displaystyle m^2}-1)^{3/2}}
{\displaystyle 2-\frac{\displaystyle M^2}{\displaystyle m^2}}}=
\frac{1+2p}n
 \end {equation}
and
\begin{equation}
\label{m}
m=
{\displaystyle 
\frac
{\sqrt2\sqrt{1+2p}}
{\sqrt{\displaystyle \frac{\displaystyle M^2}{\displaystyle m^2}-1}}}
m_{pl}
\end{equation}
For $p\gg |n|$ we have
\begin{equation}
\label{spectr1}
m\approx
2\sqrt{p}\quad m_{pl}
\end{equation}
This corresponds to the shells of large mass whose mean value radius is 
rather close to the horizon.
In the opposite case , $p\ll n$,
\begin{equation}
\label{spectr2}
m\approx
\sqrt{2}(1+2p)^{1/6}n^{1/3} m_{pl}
\end{equation}
This corresponds to the massive shells with mean radius very far from horizon.

This the end of this Section we would like to consider the behavior of the
solutions in the vicinity of the horizon (sub-Planckian deviation ), where
$|y|\gg \zeta $ $(s\sim 1)$. To be specific we will be interested in the 
solutions $R_+$ and $T_-$ regions. The expansion (\ref{expansion}) is no  more 
 valid for
the function  $F^{1/2}(s\pm i\zeta)$ but it is still valid for the wave $\Psi$.
Keeping the leading terms only we have now 
\begin{equation}
\label{s}
\Psi_{ss}(s)- \frac{2}{\displaystyle \zeta}\Psi_{s}(s)
+\left( \frac{4}{\displaystyle \alpha\zeta^{5/2}}
 (1-\frac{\displaystyle M^2}{\displaystyle 4m^2})
-\frac{2}{\displaystyle \zeta^2}
\right)\Psi(s)=0
\end{equation}
with the solution 
\begin{equation}
\Psi\sim e^{ks}, \quad k\approx - \frac{1}\zeta \pm
\sqrt{- \frac{4}{\displaystyle \alpha\zeta^{5/2}}
 (1-\frac{\displaystyle M^2}{\displaystyle 4m^2})} 
\end{equation}
The coefficient $\alpha$ equals to 1 in the $R_+$-region and to imaginary
unit i in the $T_-$-region.

In the $R_+$ region 
\begin{equation}
k\approx - \frac{1}\zeta \pm
i\sqrt{- \frac{4}{\displaystyle \zeta^{5/2}}
 (1-\frac{\displaystyle M^2}{\displaystyle 4m^2})} 
\end{equation}
and we have superposition of two waves (ingoing and out outgoing) with
relatively equal amplitudes.

in the $T_-$-region
\begin{eqnarray}
k\approx - \frac{1}\zeta \pm
\sqrt{- \frac{4i}{\displaystyle \zeta^{5/2}}
 (1-\frac{\displaystyle M^2}{\displaystyle 4m^2})} =
 - \frac{1}\zeta \pm
 \frac{\sqrt{2}}{\displaystyle \zeta^{5/4}}
 \sqrt{(1-\frac{\displaystyle M^2}{\displaystyle 4m^2})}(1+i)= \\
\nonumber
\left(\pm  \frac{\sqrt{2}}{\displaystyle \zeta^{5/4}}
 \sqrt{(1-\frac{\displaystyle M^2}{\displaystyle 4m^2})}-\frac{1}\zeta)\right)
\pm i  \frac{\sqrt{2}}{\displaystyle \zeta^{5/4}}
 \sqrt{(1-\frac{\displaystyle M^2}{\displaystyle 4m^2})}
\end{eqnarray}
The existence of two waves in the $T_-$-region reflects the quantum trembling
of the horizon . But the outgoing wave is enormously damped relative to the 
ingoing wave (of course , in the $T_-$- region the situation is exactly 
inverse one). It is this damping that cause (in a quasiclassical regime)
existence of the single ingoing wave in the $T_-$-region at the distances
larger than Planckian.

\newpage 
%%%%%%%%%%%%%%%%%%%%%%%%%%%%%%%%%%%%%%%%%%%%%%%%%%%%%%%%%%%%%%%%%%%%%%%%%%%%%%%
\section{Discussion}
%%%%%%%%%%%%%%%%%%%%%%%%%%%%%%%%%%%%%%%%%%%%%%%%%%%%%%%%%%%%%%%%%%%%%%%%%%%%%%%%
In the concluding remarks we would like to discuss the obtained results.

(1) We constructed the geometrodynamics for the spherical gravity with 
self-gravitating thin dust shell as a source. Such a shell provides us with
the only dynamical degree of freedom which is otherwise absent in pure 
spherical gravity. We  managed to separate the field canonical variables
describing the gravitational field outside the shell from the single 
pair of the shell canonical variables which are just the radius of the 
shell and the corresponding conjugate momentum. The Hamiltonian constraint
is derived for the shell which depends only on the invariants of the inner
and outer parts of the manifold and on the parameters of the shell. 
The quantum functional is subject to both quantum ADM (field) constraints
and the shell constraint. After solving the (trivial) quantum field 
constraints we are left with the functional which depends on the inner and
outer masses of the corresponding Schwarzschild manifolds and is
function  of the radius of the shell. Thus , the functional becomes a wave
function , and the remaining shell constraint is just a Schroedinger equation
for this wave function.

(2) The obtained Schroedinger equation is the equation is the equation in the 
finite differences rather than differential equation. And the shift is along  
  an imaginary axis . Dealing with the differential equations we always 
require (or assume) that the solutions should be sufficiently differentiable.
Analogously in our situation we must require solutions to be analytic 
functions except some isolated points . Our equation has branching points 
, so , the solutions will have branching points as well.

(3) The Schroedinger equation we obtained contains , as a coefficient 
function , the square root of the $F = 1 - \frac{2Gm}{R}$, which is invariant 
function of the Schwarzschild solution. This function is positive outside 
the event horizons on the both sides of the Einstein-Rosen bridge ($R_+$ and
$R_-$- regions) and it is negative beyond the horizons in the $T_+$-region
of the inevitable expansion and in the $T_-$-region of the inevitable 
contraction. We suggest to consider the square root as a function  of complex 
variable ,$(F)^{1/2}$  acquiring different phases in patches of the complete
Schwarzschild manifold. The aim of such a procedure is twofold. First, it 
allows us to obtain a common wave function covering the hole Penrose
diagram $(R_{\pm}$ and $T_{\pm}$-regions). And , second, we remove the double
cover degeneracy when the same value of radius R corresponds to two different 
points , on in the $R_+(T_+)$-region and the other is in the 
$R_-(T_-)$ -region. Now this different points with the same value of radius 
lie in different sheets of the Riemannian surface.

(4)The requirement of the analyticity of the wave function on the 
corresponding Riemannian surface means that the branching points 
should be of the same kind in order to be connected by cuts. In
another words , the number of the Riemannian sheets emerging at the
branching points must be the same.

In our case we have the branching points at  infinity and at the horizon.
But , we have two different horizons. One of them separating, say ,$R_+$
and $T_-$-regions , lies on the one sheet and the other one , separating 
$T_-$ and $R_+$-regions lies on another sheet of the Riemannian surface .
 Thus comparing the branching points at  infinity  and at the horizon in 
$R_+$-region  we obtain the first quantum number characterizing the mass
 spectrum of the system in question and, comparing these points in the $T_-$
region we obtain the second quantum number.
Thus , the mass spectrum of the black holes and wormholes should depend
on two quantum numbers .
(Note that using the same method one can obtain the famous spectra like 
oscillator, hydrogen atom and so on).

(5)It is well known that the classical theory may give rise to different 
inequivalent quantum theories. The origin of this phenomenon is a
 non-commutativity of dynamical variables and their  conjugate momenta.
The investigation of the quasiclassical limit helps to choose the ``correct''
quantum version . In our case the quasiclassical regime coincide
with the limit of large (comparing to the Planckian mass) black holes.
The finite differences equation can now be  expanded in series  with respect
to the small parameter and we can cut the series to obtain the differential
equation. We showed that  our choice of the quantum Hamiltonian give 
a good quasi-classics (ingoing wave in the $T_-$region , and outgoing 
wave in the $T_+$-region, as should be expected). Moreover we showed
that the black hole and wormhole discrete mass spectra is determined by
two quantum number  making this spectra quasicontinuous.
This resembles the appearing of the fine structure due to removing some
degeneracy (in our case it is a double covering degeneracy).
In the ordinary quantum mechanics we are used to the fact that the number
of quantum numbers equals to the numbers of degrees of freedom. From the first
sight we have  only  one dynamical degree of freedom in our problem.
Indeed, the motion of the spherically symmetric thin shell is described 
by the radius as a function of time in Lagrangian picture and by the 
radius and its conjugated momenta in the Hamiltonian picture. But actually
we have two different parts  of the Penrose diagram  with the same value of 
radius , namely $R_+$ and $R_-$-regions (and $T_+$ and $T_-$-regions).
And by our consideration they lie on different sheets of the Riemannian surface.
Of course, classical motion is forbidden in $R_-$region in the black hole case
(in $R_+$region in the wormhole case). But in the quantum theory such
motion is allowed. Thus , we have in fact two degrees of freedom . And this 
is just the origin if the second quantum number.

(6)This last remark concerns the problem of small black holes with mass about
 Planckian mass or smaller. The shift in our equation in finite differences 
is of order of the horizon size or even larger. This means that for small 
masses our equation does not feel the very existence of the horizons. And it
 gives some hope that there are no black holes at all with masses smaller than 
the Planckian mass.
%%%%%%%%%%%%%%%%%%
 \section { Acknowledgments.} 
%%%%%%%%%%%%%%%%%%
The authors are grateful to   V.A.Kuzmin, J.L.Buchbinder, 
 \    D.V.Semicoz, V.N.Melnikov for valuable discussions .

We are greatly indebted to the Russian Foundation for Basic Researches
for the financial support (Grant N.97-02-17-064).

One of us (V.A.B.) would like to thank the Astrophysics Department of the
Fermi National Accelerator Laboratory , and especially Rocky Kolb and Josh
Freeman , for the warm hospitality during his visit  to U.S.A, where 
the present work was started . he extends special thanks to D.O. and D.O.Ivanov.

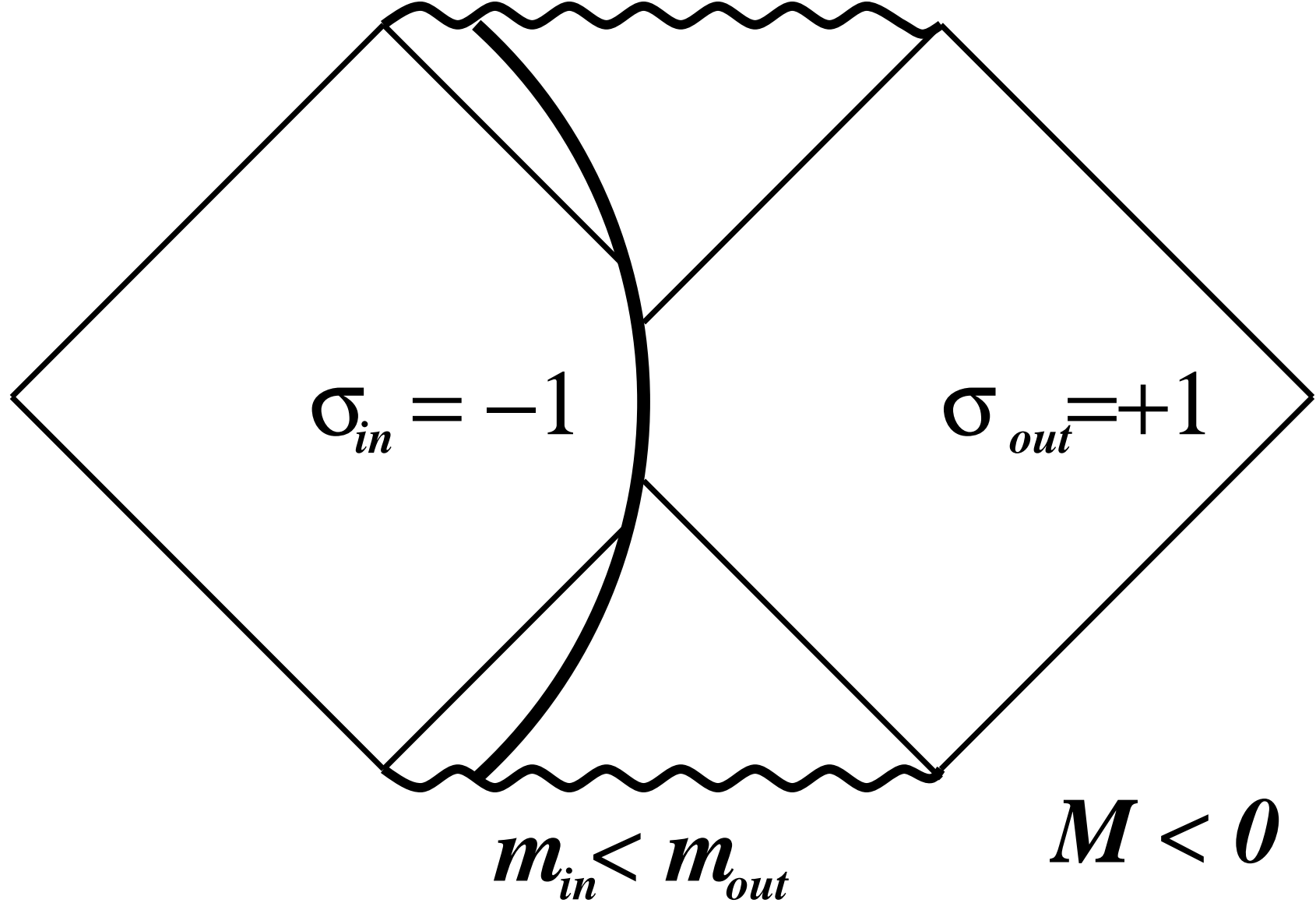
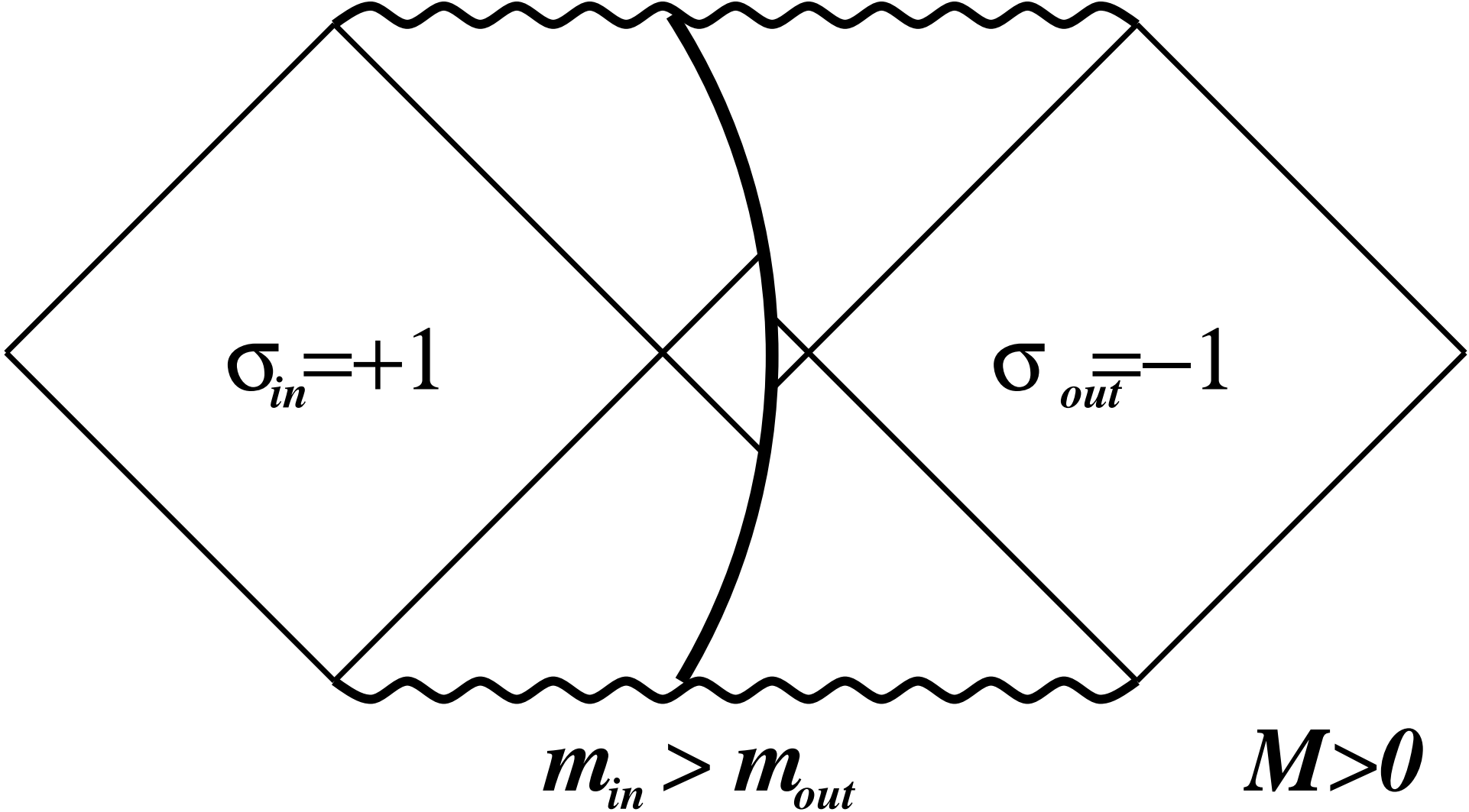
\end{document}